# The emergence of the physical world from information processing

Brian Whitworth

*Massey University, Albany, New Zealand*
*email: b.whitworth@massey.ac.nz*

*Not only is the universe stranger than we imagine, it is stranger than we can imagine*

Sir Arthur Eddington

## ABSTRACT

This paper links the conjecture that the physical world is a virtual reality to the findings of modern physics. What is usually the subject of science fiction is here proposed as a scientific theory open to empirical evaluation. We know from physics how the world behaves, and from computing how information behaves, so whether the physical world arises from ongoing information processing is a question science can evaluate. A prima facie case for the virtual reality conjecture is presented. If a photon is a pixel on a multi-dimensional grid that gives rise to space, the speed of light could reflect its refresh rate. If mass, charge and energy all arise from processing, the many conservation laws of physics could reduce to a single law of dynamic information conservation. If the universe is a virtual reality, then its big bang creation could be simply when the system was booted up. Deriving core physics from information processing could reconcile relativity and quantum theory, with the former how processing creates the space-time operating system and the latter how it creates energy and matter applications.

## 1. Introduction

We know that processing can create virtual worlds with their own time, space and objects[1], but that the physical world arises this way is normally a topic of science fiction not physics. Yet the reader is asked to keep an open mind and not reject a theory before evaluating it. The *virtual reality conjecture* is quite simply that the physical world arises from quantum processing as images arise on a computer screen. A method to evaluate this conjecture is also proposed.

---

[1] For example Second Life, http://secondlife.com/

## 1.1 Strange physics

The theories of modern physics often seem strange, e.g. in many-worlds theory each quantum choice divides the universe into alternate realities (Everett, 1957), so everything that can happen does happen somewhere in an inconceivable "multiverse' of parallel worlds. In Guth's inflationary model, our universe is just one of many possible "bubble universes" (Guth, 1998). In string theory the physical world has ten spatial dimensions, six of them "curled up" and hidden from view. M-theory puts our universe on a three dimensional "brane", floating in time on a fifth dimension we cannot see (Gribbin, 2000, p177-180). The cyclic-ekpyrotic model postulates that we are in one of two 3D worlds that collide and retreat in an eternal cycle along a hidden connecting dimension (J. Khoury, 2001).

Yet the empirical findings of physics are even stranger, e.g. the sun's gravity bends light traveling past it by "curving" nearby space. Gravity also slows down time itself, so an atomic clock atop a tall building ticks faster than one on the ground. Yet a clock in a moving plane ticks slower than one on the ground and is also heavier, as movement increases mass. Despite this malleability of space and time, the speed of light is fixed, e.g. light shone from a spaceship going at nearly the speed of light still leaves it at the speed of light. None of this makes much common sense but the experiments have been done. In 1972 one of two synchronized atomic clocks was flown in an airplane for days and another kept stationary on the ground. Less time ticked by for the moving clock. Time really does slow down with high speed travel (Hafele & Keating, 1972).

If cosmic events are strange, micro-cosmic events are even stranger. When quantum particles entangle what happens to one instantly affects the other, even if they are light years apart. The vacuum energy of "empty" space generates virtual particles with measurable effects. In Young's two slit experiment entities somehow manage to go through both slits at once, even when sent through one at a time. Quantum events like gamma radiation are entirely random, i.e. physical effects without a physical cause. Even Einstein never came to terms with quantum physics, perhaps because it makes even less common sense than relativity.

In conclusion, it isn't the theories of physics that are strange but the world itself. Physics has polled our reality and the results are in: *the physical world is stranger than it seems.*

## 1.2 The semantic vacuum

Modern physics began with Maxwell's wave equations in 1900, Einstein's special relativity in 1905, and general relativity in 1915. Despite scientific skepticism, these theories met every experimental and logical test their critics could devise. Their predictive success surprised even their advocates, e.g. in 1933 Fermi pre-discovered the neutrino before research verified it in 1953, and Dirac's equations similarly predicted anti-matter before it too was later confirmed. These and other stunning successes have made quantum mechanics and relativity theory *the crown jewels of modern physics.* They have quite simply never been shown wrong. Yet, a century later, *they still just don't make sense*. As Ford says of quantum theory:

*"Its just that the theory lacks a rationale. "How come the quantum" John Wheeler likes to ask. "If your head doesn't swim when you think about the quantum," Niels Bohr reportedly said, "you haven't understood it." And Richard Feynman … who understood quantum mechanics as deeply as anyone, wrote: "My physics students don't understand it either. That is because I don't understand it."" (Ford, 2004, p98)*

Similar statements apply to relativity theory. For perhaps the first time in the history of any science, scholars simply don't personally believe what the reigning theories of their discipline are saying. They accept them as mathematical statements that give correct answers, but not as reality descriptions of the world. This is, to say the least, an unusual state of affairs.

Relativity theory and quantum mechanics contradict not only common sense but also each other. Each works perfectly in its domain, relativity for cosmic macro-events and quantum theory for atomic micro-events, but together they clash, e.g. in relativity nothing travels faster than light but one entangled quantum entity instantly affects the other anywhere in the universe. As Greene notes:

*"The problem … is that when the equations of general relativity commingle with those of quantum mechanics, the result is disastrous." (Greene, 2004, p15)*

The problem isn't lack of use, as these theories permeate modern physics applications, from micro-computers to space exploration. By some estimates over 40% of US productivity derives from technologies based on quantum theory, including cell phones, transistors, lasers, CD players and computers. Physicists use quantum theory because it works, not because it makes sense:

*"… physicists who work with the theory every day don't really know quite what to make of it. They fill blackboards with quantum calculations and acknowledge that it is probably the most powerful, accurate, and predictive scientific theory ever developed. But … the very suggestion that it may be literally true as a description of nature is still greeted with cynicism, incomprehension, and even anger." (Vacca, 2005, p. 116)*

We have precision, proofs and applications, but not understanding. We know the mathematics exactly, but can't connect it to our experience of the world, e.g. Feynman's "sum over histories" method calculates quantum outcomes by assuming electrons simultaneously take all possible paths between two points, but how can a basic physical entity like an electron travel *all* possible paths between two points at the same time? While most theories increase understanding, such theories seem to take it away.

Despite a century of validation, neither relativity or quantum mechanics concepts are taught in high schools today, not because of their complexity, but because the emperor of modern physics has no semantic clothes. Who can teach the unbelievable? Physics has quarantined the problem behind a dense "fence" of mathematics:

*"… we have locked up quantum physics in "black boxes", which we can handle and operate without knowing what is going on inside. (Audretsch, 2004) (Preface, p x).*

Physicists use these mathematical black boxes like magic wands, but why the "spells" work we don't really know. Like monkeys in a New York apartment, we know that pressing the switch turns on the light, but not why. Pragmatists say that if the formulae work we don't need to know why, but others feel that the formulae that describe ultimate reality warrant an explanation:

*"Many physicists believe that some reason for quantum mechanics awaits discovery." (Ford, 2004, p98)*

One cannot relegate quantum and relativity effects to the "odd" corner of physics, as in many ways these theories *are* modern physics. Quantum theory rules the microcosmic world, from which the world we see emerges, and relativity rules the cosmic world that surrounds us. These two poles encompass everything we see and know of the physical world. It is unacceptable that their prime theories, however mathematically precise, remain opaque to human understanding.

Traditional objective reality concepts have had over a century to give meaning to relativity and quantum physics. That they have not yet done so suggests they never will. Hence let us now think the unthinkable alternative: *that the physical world is not an objective reality but a virtual reality*.

## 2. The virtual reality conjecture

While never commonly held, the idea that physical reality isn't the ultimate reality has a long pedigree. In Buddhism, the discriminated world is just an effect created by a universal "essence of mind" that underlies all.

In Hinduism the world of *Maya* or illusion is created by God's "play" or *Lila*. In western philosophy, Plato's cave analogy portrays the world we see as mere shadows on a cave wall that only reflect an external light[2]. The idea that the world is calculated has an equally long history. Over two thousand years ago Pythagoras considered numbers the non-material essence behind the physical world. Plato felt that "*God geometrizes*" and Gauss believed that "*God computes*" (Svozil, 2005). Both derived nature's mathematics from the divine mind, as Blake shows Urizen, "The Ancient of Days", wielding a compass to calculate the world (Figure 1). Zuse expressed the idea in modern scientific terms by suggesting that space calculates (Zuse, 1969), and since then others have explored the concept (Fredkin,

---

[2] In the analogy, people are tied up in a dark cave with their backs to its exit. Looking at the cave wall, they see only their shadows, created by sunlight from the outside, and take those shadows to be all of reality.

1990; Lloyd, 2006; Rhodes, 2001; Schmidhuber, 1997; Svozil, 2005; Tegmark, 2007; Wolfram, 2002). Some common responses to the idea are detailed in Appendix A.

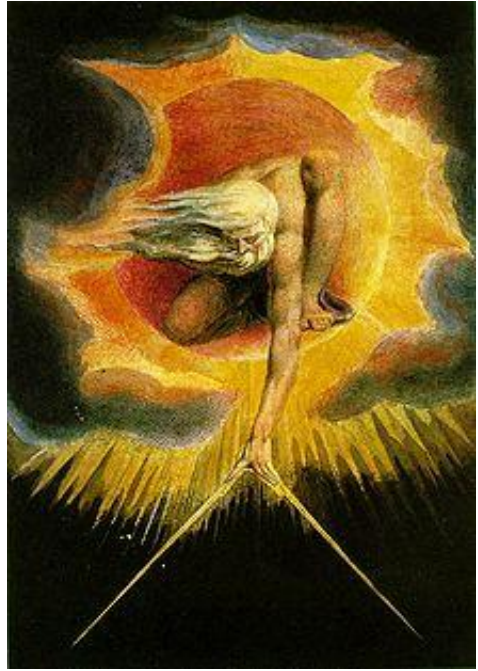

Figure 1. The Ancient of Days [3]

*2.1 Axioms of existence*

A *virtual reality* is a world created entirely by information processing, where *information* arises when a value is chosen from an available value set (Shannon & Weaver, 1949) and *processing* is the transformation of information values. As virtual worlds exist by processing, by definition nothing in them exists independently in or of itself. If the processing stops so does the virtual reality. In contrast, an *objective reality* simply *is,* and needs nothing else to sustain it. These two hypotheses are:

*1. The objective reality hypothesis:* That our reality is an objective reality that exists in and of itself and being self-contained needs nothing beside itself.

*2. The virtual reality hypothesis:* That our reality is a virtual reality that only exists by information processing beyond itself, upon which it depends.

Whatever one's personal view, these hypotheses are mutually exclusive. An objective world can't be virtual, and a virtual world can't be objective. Each theory has implications, e.g. if the physical universe is a permanent objective reality, then it has nowhere to come from or go to.

---

[3] From Wikipedia http://en.wikipedia.org/wiki/Urizen

To illustrate the contrast, consider what some call the prime axiom of physics:

1. *There is nothing outside the physical universe* (Smolin, 2001 p17).

So for example, space is assumed to have no meaning except as the relationships between real objects in the world. Yet the virtual reality conjecture turns this axiom it on its logical head:

2. *There is nothing inside the physical universe that exists of or by itself.*

This alternative axiom applies because every virtual reality must arise from processing outside itself, or its creation couldn't begin. These aren't the only statements possible about the world, but as mutually exclusive statements they provide a contrast that science can evaluate.

Philosophers have long known that one can't prove reality assumptions (Esfeld, 2004), so claims that the virtual reality conjecture cannot be tested by science to the standard of objective reality are hollow (Mullins, 2008) as science has never *proved* the world is a objective reality, either by logic or experiment. It is hypocritical to call a new theory unprovable when the established theory is in the same boat.

Science doesn't prove theories, nor test them in isolation. In practice, it merely *picks the most probable* of mutually exclusive hypotheses, here that the world is an objective reality or that it is a virtual reality. It is this *contrast,* not virtual reality theory alone, that can be tested by science.

## 2.2 The containing reality

A corollary of the virtual reality conjecture is that every virtual world must have at least one dimension outside it, in its *containing reality*. An objective reality's extra dimensions must exist inside it, so string theory's invisible extra dimensions are assumed "curled up" so small we can't see them. However in a virtual reality, invisible extra dimensions can be very large, if they exist in the containing reality. If the "extra" dimensions of physics can be inside or outside the physical world, nothing in science favors either view, as the contrast between an unknowable "in-the-world" dimension and an unknowable "out-of-the-world" one is untestable.

A common critique of the virtual reality conjecture is that it gives:

"…*no means of understanding the hardware upon which that software is running. So we have no way of understanding the real physics of reality.*"(Deutsch, 2003).

That any containing reality *must* use "hardware" like ours, or that everything real *must* be understandable to us, are just assumptions. There is no reason per se why our reality has to be the only reality, or why all reality must be knowable to us. This conjecture is not nullified because it doesn't meet the convenient and habitual assumptions of the objective reality theory it denies.

Yet the virtual reality conjecture is still a theory about *this world,* not another unknowable one. It states that *this world* is a virtual reality created by processing, not an objective reality that exists inherently by itself alone. Unprovable speculations about other virtual universes (Tegmark, 1997), or that the universe could be "saved" and "restored" (Schmidhuber, 1997), or that one virtual reality could create another (Bostrom, 2002) fall outside its scope. It certainly uses non-physical concepts, but only as other physics theories do, e.g. the quantum wave function has no counterpart in physical reality.

### 2.3 Science in a virtual reality

Science, our way of finding the truth, needs physical data to work because that is all *we can know.* This is a limitation of ourselves, not of reality. Equally, science is a way to ask questions about reality, not a set of fixed assumptions about it (Whitworth, 2007). It limits not the questions we ask but how we answer them. So to question physical reality doesn't deny science, but engages its very spirit of inquiry. Science itself is not limited to physical "observables", as it includes mathematics, and non-observables like electrons and quarks are accepted if they are *evaluated* by observation, e.g. the big bang is by definition an unobservable event, but science accepts it as true by the observable evidence of cosmic background radiation. If data from the world can decide if an unobservable big bang occurred, it can also decide if the virtual reality conjecture is true.

Conversely, could a virtual reality support science? Suppose one day the processing behind the virtual online world The Sims allowed some Sims to "think". To practice science, they would need information to test theories against. This a virtual reality could easily provide. If simulated beings in a simulated world acquired science, would they find a world like ours? Could they deduce that their world was virtual, or at least likely to be so? They couldn't *perceive* the processing creating them, but they could still *conceive* it, as we do now. Their science could then evaluate that conjecture by how their world behaved. Not only does science allow the virtual reality conjecture, a virtual reality could also allow science.

### 2.4 Local reality

In the science-fiction movie The Matrix, people lived in a virtual reality that appeared real to them as long as they stayed within it, knowing their world only by the information they received from it, as we know ours. In the story, when a pill disconnects the hero from the matrix input he falls back into the real physical world, where machines are farming people's brains in vats, i.e. the physical world is the primary reality creating the simulation. The virtual reality conjecture is the *opposite idea*: that the physical world *is* the simulation, not what creates it. It implies a quantum containing reality behind physicality, but gives it no physical properties.

Still, the usual straw man objective realists attack is Bishop Berkeley's *solipsism,* that the physical world is a hallucination, where a tree falling in a wood makes no sound if no-one is there to hear it. Dr Johnson is said to have reacted to the idea of the world as a dream by stubbing his toe on a stone and saying "*I disprove it thus*". The virtual reality conjecture is again the *opposite idea,* as it accepts that there is indeed a real world that exists apart from us. It just adds that the world we see isn't it. That the physical world is a virtual reality doesn't make it an illusion, and that the physical world is not objectively real doesn't mean that nothing is.

To clarify the difference, viewed from our physical world a simulated game world is "unreal", but to an avatar *in that world* its events are as real as it gets. Even if a virtual blow only creates virtual pain to a virtual avatar, toe stubbing will still hurt. Further, if a person is *identified* with a virtual game its events become real - imagine the identification possible in a multi-media, multi-player game with the bandwidth of our reality. A virtual world that is real within itself but still externally created can be called a *local reality.* Local and objective realities differ is not how their inhabitants see them, but in whether they need anything outside themselves to exist.

Even physical existence is relative to the observer, e.g. a table is only "solid" to us because we are made of the same atomic stuff as it is, but to an almost massless neutrino the table is a ghostly insubstantiality through which it flies, as indeed is the entire earth. *Only things constituted the same way are substantial to each other.* So in a local reality, pixels could register other pixels as "real", but still be just information patterns to the containing reality. Such a reality could look like an objective one, as Hawking says:

> "*But maybe we are all linked in to a giant computer simulation that sends a signal of pain when we send a motor signal to swing an imaginary foot at an imaginary stone. Maybe we are characters in a computer game played by aliens.*"(Vacca, 2005, p131)

Yet to give context, the next sentence was "*Joking apart, …*". For some reason even to imagine the world is virtual can only be a joke with aliens. Yet if logically the world *could* be a local reality and if physically it *behaves* like one, shouldn't we at least consider the possibility?

*2.5 The processing connection*

Physics could connect the physical world to information processing in three ways:

1. *Calculable universe hypothesis:* That processing *could* calculate physical reality.

2. *Calculating universe hypothesis*: That processing calculates *some* physical reality.

3. *Calculated universe hypothesis:* That processing calculate *all* physical reality.

*The calculable universe hypothesis* states that information processing could simulate physical reality (Tegmark, 2007). Calculable here does not mean deterministic as processing can be probabilistic, nor mathematically definable as not all definable mathematics is calculable, e.g. an infinite series. Many scientists accept that the universe is calculable in theory, as the Church-Turing thesis states that for any specifiable output there is a finite program capable of simulating it. If our universe is lawfully specifiable, even probabilistically, then in theory a program could simulate it. This hypothesis doesn't say the universe *is* a computer but that it could be simulated by one, i.e. it does not contradict objective reality.

This "thin edge of the wedge" could be falsified by a non-computable law of physics, but so far none has been found.

*The calculating universe hypothesis* states that the universe somehow uses information processing algorithms in its operations, e.g. quantum mechanical formulae. Supporters of this view include mainstream physicists like John Wheeler, whose phrase "*It from Bit*" suggests that objects ("it") somehow derive from information ("bit"). Now information processing doesn't just *model* the universe, it attempts to *explain* it (Piccinini, 2007). While a computer simulation *compares* its output to the physical world, now that processing *creates* reality is a theory about how the world actually works. The world isn't just *like* a computer, but to some degree at least, it *is* a computer. This option would be unlikely if computer simulations of physics gave no value, but they do.

*The calculated universe hypothesis* goes a step further, stating that all physical reality arises from information processing outside itself. This is the virtual reality conjecture, that the physical world is nothing but processing *output*. Supporters of this "strong" virtual reality theory are few (Fredkin, 1990), with none in mainstream physics. Due to its existential cost, it will not even be considered unless it explains what nothing else can, as this paper argues it does.

These then are the three options of information processing in physics.

## 2.6 The world is not a computer

The above three statements cumulate, as each requires the previous to be true. If the universe is not calculable it cannot calculate its operations, and if its operations can't be calculated then it can't be a calculated reality. They are also a slippery slope, as if physical reality is *calculable* then it could be *calculating*, and if it is calculating then it could be *calculated*, i.e. virtual.

Currently, the calculating universe hypothesis is presented as the best option, mid-way between the normalcy of an objective universe and the shock of a virtual one:

*"The universe is not a program running somewhere else. It is a universal computer, and there is nothing outside it."* (Kelly, 2002)

Some explicitly suggest a universal quantum computer embedded in our space-time:

*"Imagine the quantum computation embedded in space and time. Each logic gate now sites at a point in space and time, and the wires represent physical paths along which the quantum bits flow from one point to another."* (Lloyd, 1999) p172.

However processing embedded in space-time cannot create space-time, and in general no world can not create itself (Whitworth, 2010), e.g. if the physical universe is a computer with by definition nothing beyond it, *how could it begin?* An entity creating itself must already exist before it does so. That the universe computes the universe is an impossible recursion (Hofstadter, 1999). *A physical universe can no more output itself than a physical computer can print out itself.* Biological properties can evolve by bootstrapping (autopoiesis), but existence itself is not a "property" that can arise in the same way. No amount of "emergence" from nothing can create something. To argue that existence emerged from itself returns us to metaphysical mysticism.

If the physical world is the processing, *what is the output?* Or if the physical world is the output, *where is the processing?* While the brain inputs, processes and outputs information like a computer, most of the world does not (Piccinini, 2007), e.g. what "input" does the sun process and what is its "output"?

If one part of the universe outputs another, how did it all begin? Suppose string theory's hidden dimensions somehow produce the universe we see as output. If these curled-up dimensions are "in the world", the big bang that created matter, energy, space and time must also have created them. If the processing that processes the world was itself produced at the beginning, the circular creation illogicality remains. Or if the big bang didn't create these extra dimensions, by what logic are they "in the world", as they existed before its creation? If the extra dimensions of string theory are "beyond the world", then something non-physical is creating the physical, exactly as the virtual reality conjecture proposes.

The physical world can't be both processor and output because *one can't have the virtual cake and eat it too*. Either the physical world is not virtual and so not a processing output, or it is virtual and its processor is outside itself. If the physical world as a universal computer outputting itself is invalid, the three earlier options reduce to two - that the physical world is an objective reality or that it is a virtual reality.

## 2.7 Duality vs. non-duality

These considerations reflect a deep philosophical divide stretching back to the contrast between Plato's ideal forms and Aristotle's empirical pragmatism. Platonic *idealism*, that the visible physical world reflects a greater unseen world, is incompatible with Aristotelian *physicalism*, that the world we see is all there is. Logically, one of these world views *has* to be wrong.

After centuries of conflict, protagonists of science and religion agreed to a compromise, that as well as this physical reality "below", another spiritual world somehow exists apart from it. In *dualism*, developed by Descartes, the realms of mind and body *both* exist side by side, equally and separately. This compromise allowed the physicality of science to coexist with the spirituality of religion. It divided scientists into atheists who believed only in the physical world, theists who also believed in a world beyond the physical, and agnostics who didn't know what to believe.

Today, dualism seems increasingly an illogical kludge of two essentially contradictory ideas, a marriage of convenience rather than truth. How can two entirely different mind and body realities simultaneously exist, or if they do, how can independent realms of existence interact? Or if they interact, which came first causally? If conscious mind "emerges" from neuronal physics then isn't the mind created by the physical brain, and so superfluous? Or if the mind causes the body, as it does a dream, why is it constrained by the laws of physics?

Why can't I dream whatever I want, e.g. to fly anywhere, as occurs in out-of-body experiences?

The dualist view of reality, as two independent worlds in one, is currently in retreat before the simpler non-dualist view *that there is only one real world*. The scientific audience of this ideological battle has generally concluded that if there is only one world, it better be the physical one science studies. So scientists increasingly accept as "self-evident" the physicalist canon that only the physical world exists.

Yet, while rejected by both conventional science and religion, another non-dual player is still logically standing on the ideological field, namely *virtualism* (Raspanti, 2000). If *physicalism* is that only the physical world exists, and *dualism* is that another non-physical reality also exists, then *virtualism* is that only that other reality exists. It claims that the "ghostly" world of quantum physics is the actual world, and that the "solid" physical world we perceive is just an image thrown up.

*Virtualism is the non-dual converse of physicalism.* It is non-dual because, like physicalism, it asserts that there is only one world. It is the converse of physicalism because it sees the physical world that instruments register as information patterns, like an image on a screen, rather than as "things" that are real in themselves. It avoids the illogical dualism of a quantum computer creating itself, and postulates no imaginary "hardware" in a metaphysical reality beyond ours. If the physical world is virtual, it makes no sense to make physicality the yardstick of reality. In this view, the physical world is not even a drop in the universal ocean of existence, but just the wave patterns on its surface.

## 3. A prima facie case that physicality is virtual

What evidence is there that virtualism is even a possibility?

### 3.1 Initial requirements

Any processing that simulated our world would have to be:

1. *Finitely allocated*. Information as a choice from a set of options doesn't permit infinite processing, nor can a universe that began expanding a finite time ago at a finite speed be infinite. The processing needed to simulate a universe as big as ours is enormous but not inconceivable, e.g. under $10^{36}$ calculations could simulate all human history and a planet sized physical computer could do $10^{42}$ operations per second, let alone a quantum one (Bostrom, 2002).

2. *Autonomous.* Once started, it must run with no further input. While human simulations need regular data input to run, in our world such input would constitute a "miracle". As these are at best rare, this simulation must generally work without miracles.

3. *Conserved.* A system that takes no input after it starts but loses the processing it has will "run down", which our universe hasn't done for billions of years of quantum events. If matter, energy, charge, momentum and spin are all information processing, their partial conservation laws could reduce to one law of dynamic information conservation. Einstein's matter/energy equation is then just information going from one form to another.

4. *Self-registering.* System interactions must allow internal observation. While human computer simulations output to an outside viewer, we see our world from within. We register "reality" when light from the world interacts with our eyes, which are also in the world. This system must be able to consistently register itself, locally at least.

*3.2 Twelve reasons to think the physical world is a virtual reality*

One of the mysteries of our world is how every photon, electron or quark seems to just "know" what to do at each moment. Super-computers running a million-million cycles per second currently take millions of seconds (months) to simulate not just what one photon does in a million-millionth of a second, but in a million-millionth of that (Wilczek, 2008) (p113). How do these tiniest bits of the universe, with no known structures or mechanisms, make the complex choices they do? How can one photon in effect do all that processing? A later paper attributes this to information copying. Other reasons the physical world could be a virtual reality include that:

1. *It was created.* In the big bang theory our universe began as a singularity arising from "nothing" at a particular space-time event. This makes no sense for an objective reality, but every virtual reality boots up from nothing (in itself).

2. *It has a maximum speed.* In our world nothing can exceed the speed of light. While odd for an objective reality, every virtual reality has a maximum pixel transfer speed, set by the refresh rate of its screen. The speed of light could simply be a processing limit of our system.

3. *It has Planck limits.* Not only energy but also space seems quantized at Plank limits, and loop quantum gravity theory uses discrete space to avoid mathematical infinities (Smolin, 2001). An objective space has no reason to be discrete, but a virtual space must be so, as it is built entirely from discrete numbers.

4. *Tunneling occurs.* In quantum tunneling an electron suddenly appears beyond a field barrier impenetrable to it, like a coin in a perfectly sealed glass bottle suddenly appearing outside it. This is explained if quantum events are just a series of probabilistic state transitions. So reality is like a movie of still frames run quickly together, where slowing the projector gives a series of discrete pictures. A world of objects that exist inherently and continuously can't allow tunneling, but a virtual reality built from discrete probability of existence frames can.

5. *Non-local effects occur.* Quantum entanglement and wave function collapse are non-local effects that *instantly* affect quantum entities anywhere in the universe. An objective reality can't do this, but all virtual reality processing is "equidistant" to the screen, and no pixel further from its program than any other. As code can run pixels anywhere onscreen, so entangled photons could just be information objects run by the same program.

6. *Space-time is malleable*. An objective reality's space-time should be as fixed as it is, but in our world dense mass and high speeds alter time and space. This is strange in an objective reality, but if mass, movement *and* space-time arise from processing, loading one could affect another, as online videos slow down if the local server is busy. If matter uses up processing, a massive body could both dilate time and curve space. If movement uses up processing, it could shorten space and increase mass. Relativity is then just a local processing load effect.

7. *It has an uncaused cause*. Einstein never accepted that quantum choice was really random, so invoked unknown "hidden variables" to explain it, but over fifty years later none have been found. If every physical event is predicted by others, that a radioactive atom decays to emit light by pure chance, when "it decides", regardless of all prior physical events, should be impossible. Yet in a virtual world, choices random to that world can be easily generated by a processor running outside it. Indeed, a virtual world needs randomness to evolve, as it is entirely predictable without it, i.e. has zero information.

8. *Empty space is not empty*. "Empty space" is the medium that limits the speed of light, and its vacuum energy spawns the "virtual particles" of the Casimir effect. In an objective reality space is "nothing at all" and from such a nothing, nothing can come. However in a virtual reality "nothing" could be *null processing*, which can host light and spawn temporary entities.

9. *Existence can divide*. In the classic "two slit experiment" a single electron goes through both slits at once to create an interference effect with itself. In Feynman's path model "particles" simultaneously travel all possible paths between two points to pick the best one. Such effects are only possible if quantum entities exist in many places at once, which they can't do in an objective reality. Yet a virtual existence can divide up like this, as it is just information.

10. *Quantum entities are equivalent*. Every electron or quark in our world is like every other. By the quantum *indistinguishability* principle it is in impossible to mark any electron apart from another. This is odd in an objective world of things that inherently exist, but in a virtual world "objects", like electrons, are just digital symbols. If every electron in the universe is from the same code, as every "a" on this page is, they all *instantiate* the same program *class*.

11. *Complementarity*. In quantum theory simple "object" properties like position and momentum have *complementary uncertainty*, so knowing one 100% makes the other entirely uncertain. This isn't measurement "noise" but a property of reality itself. If complementary properties use the same processing, one could trade-off against another (Rhodes, 2001).

12. *It is algorithmically simple*. The algorithmic simplicity of physics is far beyond what one might reasonably expect of an objective reality:

*"The enormous usefulness of mathematics in the natural sciences is something bordering on the mysterious and there is no rational explanation for it."* (Wigner, 1960)
The laws of a virtual reality are expected to be simple if they are actually being calculated.

Perhaps individually none of the above points convince, but together they cumulate into what the courts call *circumstantial evidence*. Two properties of our world that the virtual reality conjecture explains but objective reality theory cannot are now given in more detail.

*3.3 Why did our universe begin?*

In the traditional view, the objective universe "just is", so while its parts may transform, its total is a *steady state* that always was and always will be. One doesn't expect such a universe, that is all there is, to be created in a *big bang*. Over the last century, steady state and big bang theory battled it out for supremacy on the stage of science. Steady-state proponents were respected physicists who found the idea that the entire physical universe just "popped up" out of nowhere highly unlikely. Yet since all the galaxies were expanding away from us at a known rate, one could calculate the expansion back to a source occurring about 15 billion years ago. The discovery of cosmic background radiation leftover from the big bang confirmed its reality for most physicists today.

The failure of the steady state theory removed a cornerstone of support for the view that our universe exists in and of itself. If it does exist that way, there is by definition nowhere outside itself from where it could have come in the big bang. Big bang theory neatly sidesteps questions like "What existed before the big bang?" by saying that there was no time or space before the big bang, so the question is irrelevant, i.e. it "defines away" the problem.

Yet even without *our* time or space, a universe that *began* is a dependent one, so what it depends on is a valid question. Conversely, if time and space suddenly "appeared" for no apparent reason at the big bang, could they equally suddenly disappear anytime today? If nothing in our universe comes from nothing, how can the entire universe have come from nothing? That our physical universe *arose from nothing* is not just incredible, it is inconceivable. One can state the problems simply:

1.  What caused the big bang?
2.  What caused space to start?
3.  What caused time to start?
4.  How could a "big bang" occur without time or space?
5.  How could space "start" with no time flow for the starting to occur in?
6.  How could time start somewhere if there is no "there" for it to flow in?

That the physical world *began* implies that something began it. That it came from "nothing", or somehow emerged from itself, are both highly unsatisfactory answers.

In contrast, virtual reality theory *requires* a big bang. No virtual reality has existed forever and all virtual realities initiate at a specific moment with a sudden information influx. It is a virtual reality hallmark that a single event begins its existence and its space-time. Anyone who boots up a computer begins a "big bang" that also starts up its operating system.

If the world is a self-sufficient objective reality, its space and time should be the same, i.e. exist independently of anything else. So that before the big bang there was a "no time" or a "no space" is inconceivable for an objective reality. Yet that a virtual world's time and space were started up is no surprise. Its creation was indeed from nothing *in that virtual world*, and before it there was indeed no time or space *in that virtual world*. To a virtual world observer, its origin would have all the properties of our big bang. In the virtual reality conjecture, the big bang was simply when our universe was "booted up".

This approach is distinct from current attempts to attribute *everything* to the physical world, e.g. in Zizzi's Big Wow theory consciousness somehow emerged when the inflating physical universe reached the information potential of the human brain, taken as the yardstick of consciousness (Zizzi, 2003). That machine complexity can create consciousness (Kurzweil, 1999) or that voltage changes will somehow become conscious qualia, are just the imaginations of physicalism. Super-computers are no more conscious than ordinary computers because their processing architecture is the same (Whitworth, 2009). As piling many rocks together just gives a big rock, so piling many graphic boards together into a supercomputer just gives a big machine. In this model "consciousness arises" when virtual systems by self-awareness recognize their origin. As current computing design avoids the recursive processing necessary for self-awareness, computers will not become conscious any time soon.

The big bang is now an accepted part of physics. It implies a universe created by something outside itself, a concept objective reality theory can't accommodate but virtual reality theory can. Science accepts the big bang based on data, even though we can't go back to witness it, i.e. a *conjecture about an unknowable cause was resolved by knowable world data*. If science can resolve the steady state vs. big bang hypothesis contrast, it can resolve the objective vs. virtual reality contrast. To do this we need only examine with an open mind a knowable world that:

"… *has some important and surprising things to say about itself.*" (Wilczek, 2008) (p3).

*3.4 Why is there a maximum speed?*

My interest in this area began by asking why the universe had a maximum speed? Einstein *deduced* that nothing travels faster than light from how the world behaves, but gave no structural reason for it to be so. Why can't objects just go faster and faster? What actually stops them?

*The medium of light*

If light is a wave it needs a medium to transmit it, as water waves use the medium of water. Its speed should then depend upon medium properties, like elasticity. If the medium is space, the speed of light should depend on the elasticity of space. If space is nothing it has no properties at all, let alone an elasticity, so scientists originally thought that everything must move in a *luminiferous ether,* as a fish swims in water.

However if the earth orbits the sun at 108,000 km per hour, which turns even faster around the galaxy, we must be moving through the ether (Figure 2). The ether as the medium of light is a frame of reference for it, so if we are moving through the ether in some direction, light should have different speeds in different directions. However in 1887 Michelson and Morley found that the speed of light was the same in every direction, so there could not be a physical ether.

Then Einstein showed logically that the speed of light, not the ether, was the real absolute. This left space, the medium of light, as "nothing". Some say the speed of light defines the elasticity of space, but this argues backwards, that a wave defines its medium, when really the medium must define the wave, i.e. the speed of light should conclude the argument not begin it. The nature of space should define the rate of transmission through it, but how can an empty space devoid of physical properties transmit light and limit its speed?

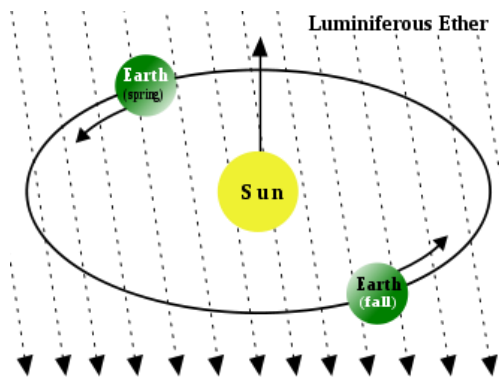

Figure 2. Luminiferous ether[4]

*The object context paradox*

An argument that physical objects need a non-physical context is as follows:

1. A world containing an inherent object must also contain something that is "not-that-object", as a boundary context.

---

[4] From Wikipedia http://en.wikipedia.org/wiki/File:AetherWind.svg

2. Unless objects entirely fill the world, the set of all objects implies a "not-any-object" context (space).

3. If space is "nothing at all", the world consists only of objects, so has no basis for movement.

4. If space exists in the world as an object, by the previous logic it also needs a context, which logic circle continues indefinitely.

For example, fish are physical objects that exist in an ocean. If the ocean is also a physical, it too needs boundary context to surround it, say land or air. If the land is also physical, it too needs a context, and so on. This circularity, of physical objects requiring physical contexts, has to stop somewhere, and in this model space is it. Yet if space exists, it can't do so as the physical objects it contains do.

*Empty space exists*

The ether error was to assume that everything must exist as physical objects do, so objects must exist in space as a fish exists in water. A physical ether isn't justified by experimental science, or by logic, as an object cannot be an ultimate context. Equally space as nothing at all contradicts much modern physics, and a world entirely of objects has no logical base for movement. The inescapable conclusion is that *empty space exists but not as physical objects do,* i.e. the medium that transmits light doesn't exist physically.

Einstein discredited the idea of a physical ether but retained the idea of physical objects. He traded Newton's old absolute space and absolute time for a new but equally absolute *space-time*:

"…*absolute space-time is as absolute for special relativity as absolute space and absolute time were for Newton …"(Greene, 2004, p51)*

He shifted the problem of how light vibrates empty space to how it vibrates an equally empty space-time, whose mathematical properties of length, breadth, depth and sequence still give no basis for media properties like elasticity. Einstein felt as strongly as Newton that *objects exist in and of themselves*, which implies an ether-like context:

"*According to the general theory of relativity space without ether is unthinkable; for in such a space there would not only be no propagation of light, but also no possibility of existence for standards of space and time …"* (Einstein, 1920, in May 5th address at the University of Leyden)

That an ether must exist but that it can't be physical led to a logical impasse he never resolved. An absolute physical reality can't have a non-physical ether around it, but a virtual reality can. If the *physical* world is virtual then the processing causing it is by definition *non-physical*. Every virtual world exists in a processing "ether" that contains its existence. So null processing can host photon calculations, as the "medium" of information waves, but still manifest as "nothing" in the virtual reality.

While physical objects existing in a physical space is illogical, virtual objects existing in a virtual space-time is not, because processing "stacks", i.e. processing can run processing, e.g. an operating system running a word-processing application is processing inside processing. Virtual objects can run in a virtual space-time if both objects and their space-time context are processing outputs. Matter and energy are then just local applications in a space-time operating system. That mass, time, space and movement all arise from processing explains not only how their object properties change, but also why they interrelate, e.g. at high speeds time dilates, space shrinks and mass increases because all involve processing. Information processing as the "quintessence" of the universe could reconcile the clash between relativity and quantum theory, with the former how processing creates space-time and the latter how it creates energy and matter.

*The speed of light is the rate of processing*

To understand this theory, an analogy with our computer processing can be useful. When a pixel moves across a computer screen, its maximum transfer rate depends on how fast the screen refreshes, e.g. a TV screen looks continuous because it refreshes faster than our eyes do[5]. In the virtual reality conjecture, a photon is just processing passed between the nodes of a universal screen that Wilczek calls *The Grid* (Wilczek, 2008). As a screen's frames per second limit how fast pixels move across it, so the grid refresh rate defines the maximum transfer rate we call the speed of light[6]. Indeed every virtual reality has a finite "speed of light" for local pixel transfers. The values we use, like 186,000 miles per second or 299,792,458 meters per second, just reflect our units. Actually, *the speed of light in a virtual reality is always just one* - one grid node per processing cycle.

In this analogy, physical matter is the pixels a screen creates. Empty space is then just a part of the screen that happens to be blank. It is "idle", as it creates no pixels, but is still "on", which null processing is the proposed vacuum energy of space. Only turning it off could show the screen (grid) itself, but that would also destroy the images on it (us). Distinguish the *pixel patterns* creating the virtual world, from the *screen nodes* processing its pixel values, and the *program calculations* that direct the processing. In this model, the physical world is the pixels, an unseen universal grid is the screen, and the equations of quantum mechanics are the programs running on it. If everything arises from an unseen quantum grid, can we "hack" into its processing? Scientists developing quantum computers may be doing just that (Lloyd, 1999).

---

[5] Usually 60 or 70 Hertz, or cycles per second is enough to look continuous to us.
[6] Given Planck time is $10^{-43}$ second, the rate would be a mind boggling $10^{43}$ hertz.

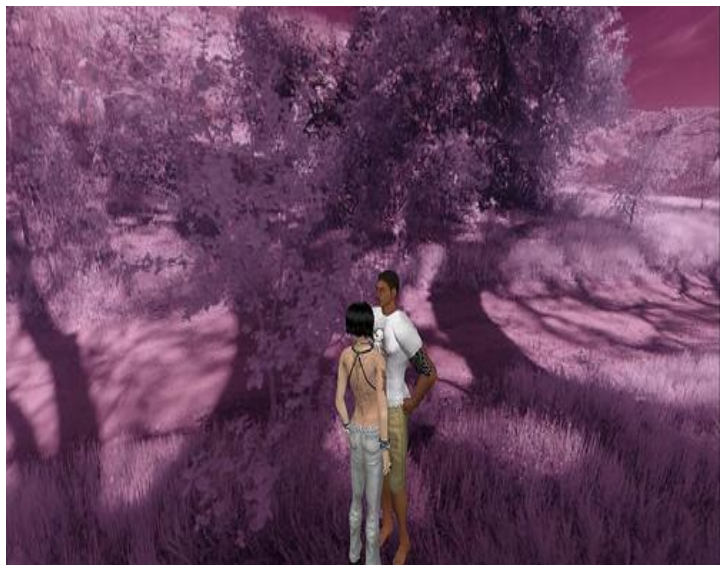

*Figure 3. A virtual reality game*

So space is not the screen the virtual reality appears on, e.g. the Figure 3 avatars are just pixels, but so is their background. As an avatar moves through the forest, no fixed node-pixel mapping is required, as any screen node can process any pixel, whether of avatar or background. Programmers "move" an avatar image through a forest by shifting all its pixels equally relative to the background, but often prefer to "bit-shift" the background behind the image, leaving the avatar centre screen as he/she "moves" through the forest. A later paper attributes relativistic frames of reference to this. Only for each processing cycle instance does a space pixel "point" necessarily map to one screen grid node. Recall that the grid proposed is not space, time, or space-time, *but what creates them*.

Processing as the ultimate context explains why transparent materials like glass slow light down, though it still goes at the maximum speed possible. If the grid that processes photon transfers also process the atoms of glass, their extra load reduces the transfer rate, making the light go more slowly. The fastest possible transfers occur when the grid has nothing else to do at all, i.e. empty space. If light passes through glass we *say* its medium is glass, and if it goes through water we *say* its medium is water, but this is just our physical bias. If water is the medium of light traveling through water, what is its medium in empty space? In this model, whether traveling in space, water or glass, the medium of light is always the unseen grid that processes everything.

That our world has a maximum speed is another accepted fact about it that virtual reality theory explains but objective reality theory cannot. What then is the "tipping point" for this case?

# 4. Evaluating the virtual reality conjecture

This prima facie case that the physical world is a virtual reality could be:

1. *Spurious*. One can satisfy any requirements by appropriate assumptions. A model can always be found to explain anything. This is less likely if the assumptions are few and reasonable.

2. *Coincidence*. The matches between virtual reality theory and modern physics are lucky coincidences. This is less likely if the matches found are many and detailed.

3. *Useful*. Seeing the world in information processing terms may open up new perspectives in physics. This response is more likely if virtual reality theory suggests new ideas.

4. *Correct*. Our world is in all likelihood a virtual reality. This is more likely if the virtual reality hypothesis explains and predicts what the objective reality alternative cannot.

How can science decide the best response?

*4.1 Method*

In science, one can't test a theory by selecting data to support it, as choosing data to fit a case is bias. So that selected computer programs (cellular automata) mimic selected world properties (Wolfram, 2002) is not evidence if the researcher *chooses* what is explained. Finding facts to fit a theory is not a new kind of science but an old kind of bias. Hence the method of this model is to *derive all core physics from information first principles*, i.e. begin with processing and derive space, time, energy and matter, explaining not just selected world events but its operational core.

This method is the usual hypothesis testing of science - assume a theory true then "follow the logic" to see if it fails, i.e. *design* then *test*. If the theory isn't true, assuming it is should soon give outcomes inconsistent with observation. If it is true, it should explain what other theories cannot.

Valid theories should be falsifiable, e.g. virtual reality theory is falsifiable as any incomputable physics would disprove it:

> *"… the hypothesis that our universe is a program running on a digital computer in another universe generates empirical predictions, and is therefore falsifiable"*(McCabe, 2004) p1

If reality does what processing can't, then the world can't be virtual, but while incomputable algorithms exist, all known physics is computable.

Objective reality theory is equally falsifiable and indeed *has been falsified*. Aspect and his colleagues showed decades ago that our world cannot be an objective reality (Aspect, Grangier, & Roger, 1982) in a well replicated experiment that challenges these assumptions of physical realism (Groblacher et al., 2007):

1. *Object locality*: That physical objects exist in a locality that limits their interactions.
2. *Object realism*: That physical objects have intrinsic properties that persist over time.

Unless formal logic itself is flawed, one or both of these objective reality assumptions must be wrong (D'Espagnat, 1979). Yet the theory remains unchallenged today, not because it is right but because no theory exists to take its place. As Chaitin, following Gödel, showed, the irreducible axioms of physics aren't logically "proven" but exist by fruitfulness - they explain more than they assume (Chaitin, 2006). Without such axioms, physics itself couldn't stand, so they are not dropped just because they are "disproved". Like house foundations, an axiom can only be removed if another can be put in to bear its load.

Foundation axioms only change during *paradigm shifts* (Kuhn, 1970), when intellectual structures are renovated and expanded, e.g. removing Euclid's axiom that parallel lines can't converge allowed the development of hyper-geometries, where parallel lines on a curved surface like the earth do converge (at the poles). Euclid's geometry is now just the zero curvature flat surface case, i.e. what was once the only possible geometry is now just one of many. If the virtual reality conjecture is also a paradigm shift, it will be evaluated by its fruitfulness not the logic of the previous paradigm, which may remain as the special case of a local reality.

# 5. Discussion

About a century ago Russell used Occam's razor[7] to cut down the idea that life is a dream:

> "*There is no logical impossibility in the supposition that the whole of life is a dream, in which we ourselves create all the objects that come before us. But although this is not logically impossible, there is no reason whatever to suppose that it is true; and it is, in fact, a less simple hypothesis, viewed as a means of accounting for the facts of our own life, than the common-sense hypothesis that there really are objects independent of us, whose action on us causes our sensations.*" (Russell, 1912)

The virtual reality conjecture is not so easily dismissed, as in physics today it is the simpler statement. Given the big bang, is it simpler that an objectively real universe arose from nothing or that a virtual reality was booted up? Given that nothing goes faster than light, is it simpler that the "nothing" of empty space limits its speed or that a processing limit does? When information processing explains more physics than common-sense (Table 1), Occam's razor cuts the other way.

## 5.1 Egocentrism

The equations of modern physics wouldn't change if the world were a virtual reality. Indeed, their status would rise, from convenient fictions to literal truths. That our physical

---

[7] Occam's Razor is not to multiply causes unnecessarily, but prefer the simpler theory

bodies are pixilated avatars in a digital world challenge not mathematics but the human ego, as science has often done before:

> *"Since our earliest ancestors admired the stars, our human egos have suffered a series of blows."* (Tegmark, 2007)

Copernicus challenged the paradigm that the Earth was the center of the universe. Science now knows that our little planet circles a mediocre star two-thirds of the way out of an average size galaxy of a million, million stars, in a universe of at least as many galaxies, i.e. we aren't the *physical* center of anything.

## Table 1. Physical outcomes and virtual causes

| Physical Outcome | Virtual Cause |
|---|---|
| *The big bang*. The universe was created from a "big bang" event that also made time and space | *Virtual reality creation*. All virtual worlds arise when an information influx starts their space-time |
| *Quantization*. Mass, energy, time and space all seem to be quantized at the Planck level | *Digitization*. Anything that arises from digital processing must be discrete |
| *Maximum speed*. Nothing in our universe can travel faster than light | *Maximum processing rate*. A screen cannot transfer pixels faster than its refresh rate |
| *Wave function collapse*. The quantum wave function collapse is a non-local effect | *Non-local effects*. Processing is "non-local" with respect to pixels on a screen |
| *Gravity and speed effects*. Near massive bodies and at high speeds space shortens and time dilates | *Processing load effects*. Processing outputs like space and time reduce with network load |
| *Physical conservation*. Physical properties like mass either conserve or equivalently transform | *Information conservation*. A stable virtual reality must conserve dynamic information |
| *Physical law simplicity*. Physical law formulae have a remarkable mathematical simplicity | *Algorithmic simplicity*. A virtual universe works best if it is easy to calculate |
| *Quantum randomness*. Quantum choice is random and unpredicted by *any* world event | *Choice creation*. A processor outside a virtual reality can create randomness in it |
| *Complementarity*. Quantum entities cant have an exact position and momentum at once | *Common processing*. Complementary properties could just use the same processing |
| *Quantum equivalence*. All quantum entities, like photons or electrons, are equivalent | *Digital equivalence*. Every digital "object" created by the same code must be equivalent. |
| *Quantum transitions*. In quantum mechanics an event is a series of state transitions | *Digital transitions*. In digital movies events are a series of picture frames |

Darwin challenged the paradigm of humanity as the pinnacle of a biology built for us. Science now knows that we only evolved about three million years ago, and that over 99.9% of all species that ever lived are now extinct, e.g. 65 million years ago the entire dinosaur class mostly died out after dominating the earth for two hundred million years, much as mammals do today. Insects and plants exceed us in biomass, are often more complex genetically, and are much more likely to survive say a nuclear disaster, i.e. we aren't the *biological* centre of anything either.

Today even the paradigm of a unitary self is challenged, as a brain "split" at its highest level, into autonomous right and left cortical hemispheres, doesn't allow for an "I" (Sperry & Gazzaniga, 1967). We don't even have the *psychological* centre we imagine we do (Whitworth, 2009).

The trend is clear: *we repeatedly imagine ourselves at the centre of things then science repeatedly finds that we aren't.* Every generation thinks it has the answers and every following one finds them wrong. Why then is *now* the end of the line of human fallacies? Is not taking *our reality as the existential centre of everything* just another egocentric assumption? And would yet another ego blow, that our physical reality is not actually "reality central", be so unexpected?

In the virtual reality conjecture, physical reality is a processing product, not something that exists in itself. The evidence presented for this view is from science not religion, e.g. the physical matter we generally take as "reality" is only 4% of the universe, with dark matter (23%) and dark energy (73%) the rest (Ford, 2004, p246). If most of the universe *isn't* the world we see, why assume that what we see is all there is? Indeed, how can a finite physical world created by a "big bang", a finite time ago, conceivably be all there is?

*5.2 The challenge of physics*

Fundamental physics is currently in a bind. On the one hand objective realism faces paradoxes it can't solve and probably never will. On the other hand the speculative mathematics of string theory is going nowhere, as it can't even manage to be wrong (Woit, 2007).

In contrast, that processing creates physicality is a logical option not yet explored, as calculated entities can be started, stopped, re-started, copied and merged in ways that "objects" can't. This is not the "brain in a vat" idea of movies like The Matrix, where a real physical world creates a false virtual one, nor the hallucinatory dream of solipsism.

Yet it is true that in a virtual world, views are only calculated as needed - if an avatar looks left a left view is created and if they look right another is shown. Everywhere one looks in a virtual world, it exists, yet the views are still only created on demand. This cracks the quantum measurement problem[8], as observing a virtual entity indeed creates (a view of) it, but raises a realism problem.

Does virtualism deny *realism*, the idea of a real world "out there". As Einstein said, surely the moon still exists if no-one observes it? If brains in vats hallucinate reality from data input, how can it all be so realistic? If no-one is looking to see if a tree falls in a forest, then no tree can fall, but what if someone looks later to see if it fell - does the system calculate a consistent history to get the current view? Did it fabricate the billions of years before mankind arrived to observe?

---

[8] The quantum measurement problem is that observing a wave function makes it take a physical state, so in quantum mechanics our observation creates reality.

In some models our consciousness is critical to quantum operations, but in this model humanity has no such central role. In it, *every* quantum interaction creates a "view", so everything is always "viewing" everything else, and everything is everywhere always "being viewed". The observer of this virtual reality is not human existence, but all existence. No tree can fall in a forest unseen, as the very ground it hits "sees" it. As there are no "gaps" in this virtual reality, so there is no view history to recapitulate.

### 5.3 The physical world as an interface

In this view, the world "out there" is a quantum one of probability waves processed by an unseen grid. The "solid" matter we see then arises from electrons and quarks, which physics knows are just quantum probabilities, i.e. information. If physical reality is just probability waves interfering, that it inherently and continuously exists in and of itself is an unnecessary belief.

Yet if the world is a virtual reality, who is the player? In our single player games, virtual worlds respond consistently to one person according to how they were designed. The knower and the known are in separate realities, the one virtual and the other containing it. Players interact in the virtual world using an avatar, then log off to return to a containing world. Multiplayer games allow more realistic worlds, because their responses increasingly come from other players. This model takes that principle to the extreme, attributing physical realism to quantum "players" in the most massive multi-player simulation conceivable.

Figure 4 gives the reality model options. The first is a simple objective reality that observes itself (Figure 4a). This gives the illogicality of a thing creating itself and doesn't explain the strangeness of modern physics, but it is accepted by most people.

The second option argues that since all human perceptions arise from neural information signals, our reality could be a virtual one, which in fiction stories is created by gods, aliens or machines, for study, amusement or profit (Figure 4b). This is not in fact illogical and explains some inexplicable physics, but few people believe that the world is an illusion created by our minds. Rather they believe that there is a real world out there, that exists whether we see it or not.

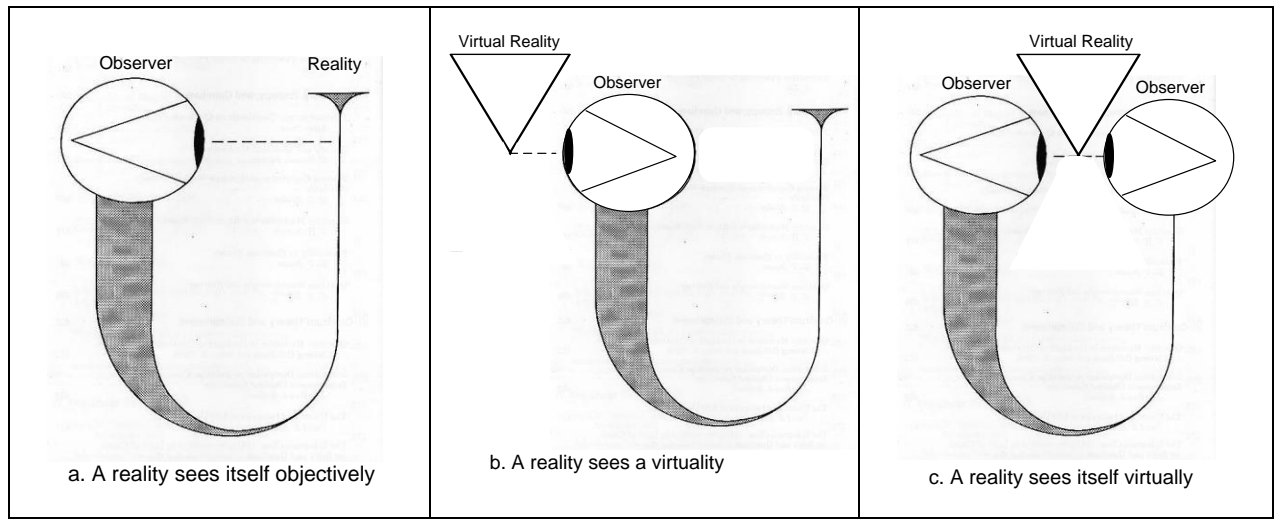

Figure 4. Reality models

The third option, of a reality that uses a virtual reality to know itself, is this model (Figure 4c). As this paper asserts and later papers expand, it is logically consistent, supports realism and fits the facts of modern physics. In it, the observer exists as a source of consciousness, the observed also exist as a source of realism, but the observer-observed interactions are equivalent to virtual images that are only locally real. This is not a virtuality created by a reality apart, but by a reality to and from itself. If the physical world is an interface to allow an existence to interact with itself, then it is like no information interface that we know.

## Appendix A. Common Responses

Common responses to the virtual reality conjecture include that it:

1) *Is just meta-physics.* Meta-physical speculation is untestable ideas about unknowable entities outside the observed world, like the number of angels on a pinhead. In contrast, the virtual reality conjecture is a hypothesis about *this world,* albeit that meta-physics (outside the world) causes physics (the world).

2) *Can't be proved.* True, but objective reality theory isn't "proved" either. Science doesn't prove theories absolutely - it just rejects improbable ones. In modern physics, it seems increasingly unlikely that the world is an objective reality.

3) *Postulates the unseen.* Being perceivable is not a demand of science or one could argue that since we can't see atoms they don't exist:

4) "*Atomism began life as a philosophical idea that would fail virtually every contemporary test of what should be regarded as 'scientific'; yet, eventually, it became the cornerstone of physical science.*" [12] p3

5) *Contradicts Occam's razor.* Occam's razor is not to multiply causes unnecessarily, to take the simplest theory that fits the facts. A hundred years ago it favored a common sense view of the world as an objective reality. Today virtual particles seethe from empty space, quantum objects teleport past impassable barriers and space and time bend and dilate. Now virtual reality theory is the simpler explanation, i.e. *Occam's razor cuts the other way*.

6) *Means the world is fake*. A virtual world need not be a fake world. The virtual reality model contradicts physical realism but not *philosophical realism:* that there is a real world "out there" generating experiences. A virtual world can be real to its participants, i.e. *locally real*.

7) *Contradicts common sense*. Common sense one told us that the sun rose and set across the earth. The same senses that tell us the earth is flat also tell that it is objectively real, but common sense no longer mandates truth.

8) *Equations are enough*. Equations without understanding are not enough. Certainly they work, but what do they mean? Physics cannot just *declare* meaning to be meaningless.

9) *Implies dual realities*. This theory postulates no dualism. If physical reality is entirely virtual, then it is a derived reality, not a dual reality. There is only one world, but it isn't the world we see.

10) *Is wrong because objective reality theory is true*. This circular refutation goes like this:
    a) You propose that the physical world is created by processing
    b) But processing is always based on the physical world (assumption)
    c) So everything is physical reality anyway.
    A well known British physics journal dismissed these ideas as follows:

"*The author insists on the "virtual reality" analogy, but seems to fail to notice that virtual reality as practiced on computers deals with a physical reality based on the known laws of physics which govern electronic or other computers. … Thus we are back to physics and asking ourselves which physical laws would be governing the computer that is supporting the virtual reality framework that the writer is proposing: back to first base as they say.*"

The reviewer *assumes* that only the physical world exists, then by that assumption manages to falsify the conjecture. When it was pointed out that this was circular reasoning, "disproving" a hypothesis by assuming its antithesis, the editor's reply was that you write, we decide. Logic is no grounds for editorial appeal in academic publishing (Whitworth & Friedman, 2009).

## Questions

The following questions highlight some of the issues covered:

1) Are quantum mechanics and general relativity true statements about reality?

2) Does science require an objectively real world?

3) Would a virtual reality allow science?

4) How does a local reality differ from an objective one?

5) What is the logical opposite of physicalism, that only the physical world exists?

6) Could the world be a universal computer that calculates and outputs itself?

7) In what ways does our world act like a virtual reality?

8) Could an objective reality arise from a "big bang"?

9) If light is a wave, what medium does it travel in?

10) Why cant anything ever go faster than light?

11)    Is the virtual reality conjecture testable? Is it falsifiable? Is it provable?

12) If modern physics has falsified objective reality theory, why is it still the accepted?

13) How would the mathematics of physics change if the physical world was virtual?

14) If the world is a virtual reality, who is observing it?

# Acknowledgements

Thanks to Onofrio Russo, NJIT, for arousing my interest, to Ken Hawick, Massey University, for listening to my ramblings, to Cris Calude, Auckland University, for a useful critique, and to Jonathan Dickau, Matthew Raspanti, Bruce Maier, Tom Cambell, Ross Rhodes, Bryan Warner, Andrew Eaglen and Andrew Thomas for comments. Especial thanks to my son Alex, who helped me think and express more clearly. Still, all the mistakes are mine alone. The chapter version is can be downloaded at http://brianwhitworth.com/BW-VRT1.pdf